\documentstyle[aps,epsf,twocolumn,prl]{revtex}

\def\be{\begin{equation}}
\def\ee{\end{equation}}
\def\ba{\begin{eqnarray}}
\def\ea{\end{eqnarray}}
\def\fun#1#2{\lower3.6pt\vbox{\baselineskip0pt\lineskip.9pt
        \ialign{$\mathsurround=0pt#1\hfill##\hfil$\crcr#2\crcr\sim\crcr}}}
\def\la{\mathrel{\mathpalette\fun <}}
\def\ga{\mathrel{\mathpalette\fun >}}
\def\mpl{{{M_{{\rm Pl}}}}}

\def\prl#1#2#3{Phys. Rev. Lett. {\bf #1}, #2 (#3)}
\def\prd#1#2#3{Phys. Rev. D {\bf #1}, #2 (#3)}
\def\plb#1#2#3{Phys. Lett. {\bf #1B}, #2 (#3)}
\def\npb#1#2#3{Nucl. Phys. {\bf B#1}, #2 (#3)}
\def\apj#1#2#3{Astrophys. J. {\bf #1}, #2 (#3)}
\begin{document}
\draft

\twocolumn[\hsize\textwidth\columnwidth\hsize\csname @twocolumnfalse\endcsname
\title{Natural Candidates for Superheavy Dark Matter in String and $M$ 
Theory}
\author{ Karim Benakli $^{(1)}$, John Ellis$^{(2)}$  and Dimitri V.
Nanopoulos$^{(1,3)}$}
\address{$^{(1)}$ {\it Center for Theoretical Physics, Dept. of Physics,
Texas A \& M University, College Station, TX~77843-4242, USA }}
\address{$^{(2)}${\it Theory Division, CERN, CH-1211 Geneva 23, Switzerland}}
\address{$^{(3)}$ {\it Astroparticle Physics Group, HARC, Mitchell
Campus, The Woodlands, TX~77381, USA} and {\it Chair of Theoretical
Physics, Academy of Athens, 28~Panepistimiou~Ave.,
Athens~GR-10679, Greece}}

\maketitle
\widetext

\begin{abstract}
\noindent
We reconsider superheavy dark matter candidates in string and
$M$ theory, in view of the possibility that inflation might generate
superheavy
particles with an abundance close to that required for a near-critical
Universe. 
We argue that cryptons - stable or metastable bound states of matter in
the hidden sector -
are favoured over other possible candidates in
string or $M$ theory, such as the Kaluza-Klein states associated with
extra dimensions. We exhibit a specific string model that predicts
cryptons as hidden-sector bound states weighing $\sim 10^{12}$ GeV,
and discuss their astrophysical observability.
\end{abstract}
]

\narrowtext  There is striking evidence from
astronomical observations for the existence of dark matter. The rotational
velocities of  galaxies, the dynamics of galaxy clusters 
and theories of structure formation suggest
that most of the matter in the Universe is invisible
and largely composed of non-baryonic particles. Many candidates have
been proposed  as constituents of this particle dark matter.  
An upper bound on the self-annihilation cross section,
based on unitarity, suggests that particles much heavier than 1~TeV $\sim
\sqrt{\mpl \times T_{CMBR}}$ that have been in
thermal equilibrium would be left with such a large relic abundance
that they would overclose the Universe~\cite{griestkam}. 
This rule could be evaded if there was significant entropy generation
after they went
out of thermal equilibrium. This possibility was raised in
connection with one class of superheavy dark matter candidates:
{\it cryptons}~\cite{eln1}, which are stable or metastable bound states of
matter in a hidden sector of string theory.
However, there was until recently little
reason to expect that the abundance of cryptons or other superheavy relics
would be such as to constitute most of the mass density of dark matter.

The question of abundance of superheavy relics has recently been 
revisited~\cite{KRT,CKR,kuz1}. In particular, a
gravitational mechanism was suggested~\cite{CKR,kuz1}
whereby cosmological inflation may generate a
desirable abundance of such massive and weakly interacting massive
relic particles. Numerical analysis indicates that the 
process may be largely independent of details of the models considered
for most properties of the dark matter constituent, as well as of details
of the transition between 
the inflationary phase and the subsequent thermal radiation-dominated
phase.
In the light of this new proposal, it is interesting to revisit the 
possibility that cryptons or other superheavy string
relics may constitute an 
important part of the astrophysical dark matter.

We argue below that the combination of
metastability and a desirable relic density are more
likely for cryptons than for superheavy
Kaluza-Klein states associated with the extra dimensions compactified
at short distance scales in string or $M$ theory. We also discuss the
possible observability of cryptons via their high-energy decay products.

We first review some basic
facts about superheavy dark matter and the 
new proposals~\cite{CKR,kuz1}
for relic production during inflation. To be of
interest in the Universe today, any such
relic particle $X$ should have a lifetime at least of the order of the age 
of the Universe: $\tau_X \ga 10^{10}$~y, and there may be more
stringent limits coming from searches for its decay products,
that depend on its mass and decay modes~\cite{eglns}. One should
also
require that its energy density $\rho_X(t_0)$ at the present
time $t_0$ does not overclose the universe, i.e., $\Omega_X\equiv
\rho_X(t_0)/\rho_C(t_0) <1$, where 
the critical density $\rho_C(t_0)=3 H_0^2\mpl^2/8\pi$, with
$H_0$ the Hubble expansion rate today. 

If the particle $X$ was in thermal equilibrium with the
primordial  plasma, the expected number of particles
today $n_0$ is inversely proportional to the
self-annihilation cross section $\sigma_A$. If the latter
is bounded above by ${\cal O}(1/M_X^2)$,
as suggested by unitarity in a finite number of partial waves, then
one reaches the usual conclusion that
dark matter particle masses are unlikely to exceed 
greatly the electroweak scale. However, the dark
matter candidate might avoid this constraint if it has never reached
local thermal equilibrium with the primordial plasma. This situation is 
realized if $n_{X} \langle \sigma_A |v| \rangle \la H$,
where $n_X$ is the (conserved) comoving number density and
$|v|$ is the M{\o}ller speed for the dark matter particles
$X$. Given the usual dependence of  $\sigma_A$ on 
the mass and coupling of $X$,
the absence of thermal equilibrium requires that  i)
$X$ is weakly interacting,  and ii) $ M_X$ is of
the order of the Hubble constant.

However, without the constraint of thermal equilibrium,  
how does one explain why  the superheavy
particle is naturally generated in the correct number to form the cold dark
matter of the universe? This problem was addressed by the authors
of~\cite{CKR,kuz1}.  They suggested that these particles might be created through
the interaction of the vacuum with the gravitational field during the
reheating period of the universe~\cite{bd}. Such a process involves only
the gravitational  interactions of the particle, and thus is quite
independent of the other (weak) interactions that it might have. This
scenario leads to the following mass density of the particle $X$ created
at time $t=t_e$~\cite{CKR}:

\begin{equation}
 \Omega_X h^2 \approx \Omega_R h^2\:
\left(\frac{T_{RH}}{T_0}\right)\:
\frac{8 \pi}{3} \left(\frac{M_X}{\mpl}\right)\:
\frac{n_X(t_{e})}{\mpl H^2(t_{e})}.
\label{eq:omegachi}
\end{equation}
where  $\Omega_R h^2 \approx 4.31 \times 10^{-5}$ is the fraction of
the critical energy density that is in radiation today, and $T_{RH}$ is the 
reheating temperature.

The numerical analysis of~\cite{CKR} indicates that the correct magnitude
for the
abundance of the $X$ particle is obtained if its mass lies in the
region $0.04 \la M_X/H \la 2$, where $H \sim 10^{13}$~GeV is the Hubble
expansion rate at the end of inflation, which is 
expected to be of the same order as the mass of
the inflaton. For our purposes, 
we shall consider the range $10^{11}$~GeV $\la\,M_X\, \la \, 10^{14}$~GeV 
to be favourable for superheavy dark matter, 
and consider next various candidates for the $X$ particle 
within the context of string and $M$ theory. 
 
String theories have historically been analyzed in the weak-coupling limit,
where there is an observable sector containing the known gauge interactions
and matter particles, and a hidden sector that is expected to become
strongly interacting and may play a r\^ole in supersymmetry breaking.
In addition to the states that are massless before this and 
subsequent stages of
symmetry breaking, such string models also contain Kaluza-Klein
excitations with masses related to the scales at which surplus dimensions
are compactified. In the weak-coupling limit, all these states would have masses
comparable to the Planck mass $\mpl \sim 10^{19}$~GeV, beyond the range
favoured by~\cite{CKR,kuz1}. However, the string mass estimate may be
revised
downwards in the strong-coupling limit described by $M$ theory, requiring a
revised discussion as provided below. We now discuss in more detail
some specific string and $M$ theory possibilities.

{\it i) Level-one heterotic-string models:}
These have been the most studied vacua of string theory. The possibility of
building explicit models and carrying out detailed computations makes
possible a precise analysis. A well-established prediction of this class
of compactifications is the
existence of light ( massless at the string scale) 
states which are singlets under $SU(3)_c$ and carry 
fractional electrical charges, that appear generically in the hidden
gauge-group sector~\cite{FCP,FCP2}. Such particles cannot be free, because
the lightest of these particles would have to be stable and 
present in the Universe with a large abundance. There are very stringent
upper limits on the abundance of such a fractionally-charged relic, from
successors of the Milliken experiments, which are
many orders of magnitude below the critical density. However, 
theoretical expectations
for their abundance on Earth are about ten orders of magnitude above these
limits~\cite{FCP3}. Thus
the only viable string vacua are those where these charges
are confined by a ``hidden" group $G$, as in QCD.
The integer-charged lightest singlet bound states of such a 
hidden-sector group may be stable or metastable, providing the dark-matter
candidates termed {\it cryptons}~\cite{eln2}.

The confining group $G$ must be such that singlet bound states of
$SU(3)\times G$ have integer electric charges. For
$\displaystyle{G=\prod_N SU(N)\times\prod_n SO(2n)}$, this condition
states that~\cite{FCP2,eln2}:
\begin{equation} 
\sum_N{{i_N (N-i_N)}\over{2N}}+\sum_n \cases{0 &for $j_n =0$\cr 1/2
&for $j_n =2$\cr n/8 &for $j_n =1$\cr} 
\label{cqc} 
\end{equation}  
must be a non-vanishing integer, where for every $N$, $i_N$ is some integer
between 0 and $N-1$. Thus the electric charge of a state transforming
in the representation $N$ or $\overline{N}$ of $SU(N)$ and/or $2n$ of
$SO(2n)$ must be: 
\begin{equation}
 q= \pm\sum_N{i_N \over N}+\sum_n {j_n \over 2}~~{\rm mod}1, 
\end{equation}
with $\pm$ corresponding to representations $N$ or $\overline{N}$.

The case where $G$ is a product of semi-simple factors presents the
advantage, compared to a large unique semi-simple group, of generally
giving rise to a smaller number of fractionally-charged states that have
to be included in the running of the 
supersymmetric Standard-Model gauge couplings~\cite{BL,AB}.
Note also that, because of these states, 
the $G$ gauge sector is not completely ``hidden". This may even be
advantageous, if supersymmetry is broken when
the coupling of $G$ becomes strong, and an $F$ term is generated. This 
supersymmetry breaking would be mediated to the observable sector  not only
by gravitational interactions involving the graviton supermultiplet,
but also through the usual
Standard-Model gauge interactions via the supermultiplets of 
fractionally-charged states\cite{AB}.

{\it ii) Higher-level string models:}
This class of constructions is largely motivated by the need to
accommodate adjoint Higgs representations in GUTs other than
flipped $SU(5)$. In so doing, they usually also 
lead to new exotic matter. In the specific
case of standard $SU(5)$ realized at level 2, there were found
representations transforming as  $(8,1,0)$, $(1,3,0)$ and $(1,1,0)$ of
$SU(3)\times SU(2)\times U(1)$~\cite{sgut}. These particles 
were also found to have no superpotential, and hence should appear
below the string scale. It was suggested in~\cite{BFY} that these states 
might have masses of the order of
$10^{13}$ GeV, so as to resolve apparent discrepancy between the
unification and string scales. In such a scenario, the singlet $(1,1,0)$ 
state could be a dark matter candidate.  
In common with the level-one constructions, these models also have a
hidden sector where supersymmetry breaking might originate, which might also
provide a stable bound state as discussed above.

{\it iii) $M$ theory on $S^1/Z_2$:}
The first phenomenological studies of $M$ theory 
have raised the possibility that six
of the original eleven dimensions
might be compactified at an energy scale comparable to the
conventional supersymmetric GUT scale,
in the range of $10^{16}$ to $10^{17}$~GeV, 
leaving an effectively five-dimensional low-energy theory.
The fifth dimension would subsequently be compactified
down to four dimensions,
on an $S^1/Z_2$ segment with a size $\rho$ that might be of order
$10^{-13}$ to $10^{-15}{\rm GeV}^{-1}$~\cite{witt}. {\it ``Hexon"}
Kaluza-Klein states associated with the $11 \rightarrow 5$
compactification are likely to be too heavy to have been
produced copiously via the mechanism of~\cite{CKR,kuz1}, but
massive states associated with the $5 \rightarrow 4$
compactification might well have masses in the favored range.~\cite{foot}
We term such
Kaluza-Klein states ``{\it pentons}''.
In this type of $M$-theory scenario, the effective field theory
at energies below about $10^{16}$~GeV is $N=2$ supergravity,
which in the specific case of $M$ theory
compactified on a Calabi-Yau manifold has 
has been shown~\cite{ant} to contain $h_{1,1}-1$ vector
hypermultiplets
and $h_{2,1}+1$ scalar hypermultiplets. None of these carry conventional
gauge interactions, which are realized on the walls.

The {\it pentons}, i.e. the massive Kaluza-Klein bulk
excitations of this particular
five-dimensional supergravity theory, therefore do not 
carry any new conserved quantum number,
in contrast to generic ``pyrgons"~\cite{pyg}.
These states are free to decay into fields living 
on one of the walls of universe, and the most stable {\it pentons}
would be those with the smallest couplings to boundary fields. We expect
these to be of the order 
of the gravitational constant. Thus their lifetimes should
be of order
$\rho^3 G_N$, which is much too short to constitute the dark matter in
the Universe today.

A more likely candidate might emerge among the states living 
on the wall at the (hidden) opposite wall of the five-dimensional
bulk from the conventional observable sector. As in the previous
discussion of weakly-coupled string models, one would expect that
the mass of the lightest {\it crypton} bound state
would have its origin in non-perturbative 
strong-coupling effects, which might well
occur at the favored scale of order $10^{13}$ GeV~\cite{CKR,kuz1}.
Similarly to protons on our (observable) wall, 
there would in general be some flavour symmetry of the hidden world which
allows  
such a bound state  to remain metastable, providing us with a candidate for 
the dark matter particle $X$.

We now review an explicit example of a string model whose hidden
sector contains such metastable {\it crypton} bound states. This
model was originally constructed in the weak-coupling limit~\cite{fsu},
but
we expect that it may be elevated to an authentic $M$-theory model in the
strong-coupling limit.
This model has the gauge group $SU(5)\times U(1)\times U(1)^4\times 
SO(10)\times SU(4)$, with the latter two factors yielding
strong hidden-sector interactions. The following Table lists
the matter content of this hidden sector.

\vskip 0.3cm
{\centering \begin{tabular}{|c|c|} \hline
$\Delta^0_1(0,1,6,0,-\frac{1}{2},\frac{1}{2},0)$&
$\Delta^0_2(0,1,6,-\frac{1}{2},0,\frac{1}{2},0)$\\
$\Delta^0_3(0,1,6,-\frac{1}{2},-\frac{1}{2},0,\frac{1}{2})$&
$\Delta^0_4(0,1,6,0,-\frac{1}{2},\frac{1}{2},0)$\\
$\Delta^0_5(0,1,6,\frac{1}{2},0,-\frac{1}{2},0)$& \\
$T^0_1(10,1,0,-\frac{1}{2},\frac{1}{2},0)$ &
$T^0_2(10,1,-\frac{1}{2},0,\frac{1}{2},0)$ \\
$T^0_3(10,1,-\frac{1}{2},-\frac{1}{2},0,\frac{1}{2})$ &
$T^0_4(10,1,0,\frac{1}{2},-\frac{1}{2},0)$ \\
$T^0_5(10,1,-\frac{1}{2},0,\frac{1}{2},0)$&   \\ \hline 
\end{tabular}\par}

\vspace*{0.3 cm}

{\centering \begin{tabular}{|c|c|} \hline ${\tilde
F}^{+\frac{1}{2}}_1(1,4,-\frac{1}{4},\frac{1}{4},-\frac{1}{4},\frac{1}{2})$
& ${\tilde
F^{+\frac{1}{2}}}_2(1,4,-\frac{1}{4},\frac{1}{4},-\frac{1}{4},-\frac{1}{2})$
\\ ${\tilde
F}^{-\frac{1}{2}}_3(1,4,\frac{1}{4},\frac{1}{4},-\frac{1}{4},\frac{1}{2})$
& ${\tilde
F}^{+\frac{1}{2}}_4(1,4,\frac{1}{4},-\frac{1}{4},-\frac{1}{4},6-\frac{1}{2})$
\\ ${\tilde
F}^{+\frac{1}{2}}_5(1,4,-\frac{1}{4},\frac{3}{4},-\frac{1}{4},0)$ &
${\tilde
F}^{+\frac{1}{2}}_6(1,4,-\frac{1}{4},\frac{1}{4},-\frac{1}{4},
-\frac{1}{2})$\\ ${\tilde {\bar
F}}^{-\frac{1}{2}}_1(1,4,-\frac{1}{4},\frac{1}{4},\frac{1}{4},\frac{1}{2})$
& ${\tilde {\bar
F}}^{-\frac{1}{2}}_2(1,4,-\frac{1}{4},\frac{1}{4},\frac{1}{4},-\frac{1}{2})$
\\ ${\tilde {\bar
F}}^{+\frac{1}{2}}_3(1,4,-\frac{1}{4},-\frac{1}{4},\frac{1}{4},-\frac{1}{2})$
& ${\tilde {\bar
F}}^{-\frac{1}{2}}_4(1,4,-\frac{1}{4},\frac{1}{4},\frac{1}{4},
-\frac{1}{2})$ \\ ${\tilde {\bar
F}}^{-\frac{1}{2}}_5(1,4,-\frac{3}{4},\frac{1}{4},-\frac{1}{4},0)$ &
${\tilde {\bar
F}}^{-\frac{1}{2}}_6(1,4,\frac{1}{4},-\frac{1}{4},\frac{1}{4},
-\frac{1}{2})$\\ \hline
\end{tabular}\par}

\vspace*{0.3 cm}
Table:  {\it The spectrum of hidden matter fields 
that are massless at the string
scale in the revamped flipped $SU(5)$ model.
We display the quantum numbers under the hidden 
gauge group $SO(10) \times
SO(6) \times U(1)^4$, and subscripts indicate the electric charges.}
{}\\
 
Analysis of the calculable superpotential in this model 
shows that most of these fields acquire
heavy masses just below the string scale from
couplings with singlet fields that acquire vacuum expectation values to
cancel the $D$-term of the anomalous $U(1)$. The only light states 
that survive to have lower masses
are the $T_3, \Delta_3, {\tilde F}_{3,5}$  and
${\tilde {\bar F}}_{3,5}$. Analysis of the
renormalization-group $\beta$ functions of $SO(10)$ and $SO(6)$ suggest
that their confinement scales might lie at $\Lambda_{10}\sim
10^{14-15}$GeV for $SO(10)$ and  $\Lambda_{4}\sim 10^{11-12}$GeV for
$SU(4)$.  This indicates that the states in the $SU(4)$ representations
$\Delta_3, {\tilde F}_{3,5}$ and ${\tilde {\bar F}}_{3,5}$ will form
the lightest bound states.

In addition to meson and baryon bound states as in QCD, 
one expects quadrilinear {\it tetron} bound states
specific to $SU(4)$~\cite{eln1}.
The mesons comprise $T_iT_j$, $\Delta_i \Delta_j$ and $
{\tilde  F_i} {\tilde {\bar F}_j}$ bound states, which are 
all short-lived, as they
decay through order $N=$ 3, 4 or 6 non-renormalizable  operators. The
baryons have the
constituents $ {\tilde F_i} {\tilde F_j} \Delta_k$ and  $
{\tilde {\bar F}_i} {\tilde {\bar F}_j} \Delta_k$ are also short-lived. Finally, there are
{\it tetrons} composed of four $\tilde F_i$s, of which the  lightest
have the forms  $ {\tilde F_i} {\tilde F_j}
{\tilde F_k} {\tilde F_l} $  and  $ {\tilde {\bar F}_i} {\tilde {\bar
F}_j} {\tilde {\bar F}_k} {\tilde {\bar F}_l} $, where $i,j,k,l = 3,5$.
As in the case of QCD 
pions, one may expect the charged states~\cite{CHAMPs} to be slightly
heavier than the
neutral ones, due to electromagnetic energy mass
splitting. No non-renormalizable interaction capable of enabling
this lightest bound state to decay has been 
found in a search up to eighth order.
We therefore consider that this lightest neutral tetron
is a perfect candidate for a superheavy dark matter particle. 
A rough lower bound on the lifetime of this lightest tetron is
of the order: 

\begin{equation}
\tau_X \sim  \frac{1}{M_X} {\left(  \frac{m_k}{M_X} \right) }^{10},
\end{equation}
which is very sensitive to $M_X$ and  the scale $m_k$ of suppression of 
the non-renormalizable terms. For $m_k \sim 10^{17-18}$~GeV, 
and a tetron mass $M_X \sim 
10^{12}$GeV, we find that $\tau_X > 10^{7-17}$ years. This is a lower
bound,
and the actual lifetime may well be considerably longer if the leading 
decay interaction is of significantly higher order. 

If cryptons are unstable, their decays might explain the
ultra-high-energy cosmic ray events~\cite{uhcr} 
observed beyond the GZK cut-off
energy~\cite{gzk}. Explanations of these events as decays of
superheavy relic particle has been considered by many
authors~\cite{kuz1,kuz2,ber}. If the $X$ particles form the
same proportions of dark matter in the galaxy and in intergalactic
space, it was argued in~\cite{ber} that the observed extensive air
showers are mainly due to gamma rays and nucleons produced
by $X$ particle decays in the halo of our galaxy. This
explains the absence of the GZK cut-off, and one easily avoids
constraints on the cascade radiation. 

The observed fluxes constrain $\Omega_X /\tau_X$.
For $\tau_X\sim 10^{10}$~years, the density of $X$ particles is
$\Omega_X \sim 10^{-21}$. For the range  $\Omega_X \sim 1$ of
interest to us, we could match the observed fluxes if  
$10^{15}$~years~$\la \tau_X \la 10^{22}$~years~\cite{eglns,kuz1}.  In
contrast to other
superheavy particles that decay via non-perturbative
gauge (instanton) or quantum-gravity (wormhole) interactions, our
candidate has the advantage that decays via
non-renormalizable operators that might be explicitly computed in a
perturbative framework. The desired value might be  attainable  in the
explicit model discussed above if, for example, $m_k \sim 10^{17}$~GeV 
and the first decay interaction is of order $N=9$ or higher.

As discussed in~\cite{kuz1},  the ultra-high-energy cosmic rays
produced by the crypton decay might present a signature that
distinguish them from other possible sources: the  cosmic-ray spectrum
should have a new cut-off at an energy $\la M_X$. 

To conclude: we have re-evaluated possible candidates for superheavy dark
matter in
string or $M$ theory, in the light of new production
estimates~\cite{CKR,kuz1}.  In particular, we have revisited the
proposal that {\it cryptons} may be an important component of  dark
matter. These particles have been shown to exist in a particular model
with the desired mass, very weak interactions and long lifetimes. The
possibility that they are generated by gravitational interactions
with the vacuum~\cite{CKR,kuz1} may answer to the long-standing
question
why their abundance  should lie in the interesting range $\Omega \sim
0.1$  to $\sim 0.9$. We have also discussed other dark-matter
candidates in the context of string and $M$ theory, such as {\it
hexons} and {\it pentons}, but these do not have the desired mass
and/or metastability, at least in the models studied so far. 

 The work of K. B. (D.V.N.) was supported  by DOE grant 
DE-FG03-95-ER-40917 (DE-FG05-91-GR-40633).


\begin{references}
\bibitem{foot}It could even be that
inflation is associated with the $5 \rightarrow 4$ transition,
but we do not pursue this possibility here.
\end{references}
\end{document}